# A scalp-EEG network-based analysis of Alzheimer's disease patients at rest

Aya Kabbara [a,b,c,d*]   Wassim El Falou [a,b]   Mohamad Khalil [a,b]   Hassan Eid [e]   Mahmoud Hassan [c,d]

[a] Azm research center in biotechnology, EDST, Lebanese University, Lebanon
[b] CRSI research center, Faculty of engineering, Lebanese University, Lebanon
[c] Université de Rennes 1, LTSI, F-35000, France
[d] INSERM, U1099, Rennes, F-35000, France
[e] Mazloum Hospital, Tripoli, Lebanon
aya.kabbara7@gmail.com

*Abstract*—Most brain disorders including Alzheimer's disease (AD) are related to alterations in the normal brain network organization and function. Exploring these network alterations using non-invasive and easy to use technique is a topic of great interest. In this paper, we collected EEG resting-state data from AD patients and healthy control subjects. Functional connectivity between scalp EEG signals was quantified using the phase locking value (PLV) for 6 frequency bands, θ (4–8 Hz), α1(8–10 Hz), α2(10–13 Hz), ß(13–30 Hz), γ(30–45 Hz), and broad band (0.2-45 Hz). To assess the differences in network properties, graph-theoretical analysis was performed. AD patients showed decrease of mean connectivity, average clustering and global efficiency in the lower alpha band. Positive correlation between the cognitive score and the extracted graph measures was obtained, suggesting that EEG could be a promising technique to derive new biomarkers of AD diagnosis.

## I. INTRODUCTION

Around 30,000 people suffered from Alzheimer's disease (AD) or related dementia in Lebanon. AD is generally characterized by a progressive decline of memory and cognitive functions [1]. Nowadays, many evidences suggest that AD is associated with alterations of large-scale brain networks due to synaptic dysfunction [2]. Therefore, the identification of the alterations occurred in the brain network topology of AD patients is a topic of great interest [3].

Moreover, in some cases, AD patients show only moderate clinical symptoms which impede their diagnosis [4], [5]. This highlights the need of objective, non-invasive and easy to use biomarkers for an early detection of the disease. In this context, resting-state networks derived from electroencephalography (EEG) may be a key feature to explore the network properties of AD.

To quantify brain network, graph theory-based analysis is usually used. In graph-theory approaches, brain networks are represented as graphs composed of nodes (electrodes or brain regions) connected by edges (functional connectivity) [6]. Once nodes and edges are identified from the neuroimaging data, network topological properties can be explored by extracting specific graph metrics. In the context of resting-state analysis of AD, numerous functional and structural studies and have been conducted [7]–[9]..

Here, we explored the topological differences between the networks of AD and healthy controls using EEG. For this end, we collected EEG data from 14 subjects at rest (eyes closed). We then reconstructed the functional networks at the scalp-level using the phase synchronization method [10] for θ, α1, α2, ß, γ, and broad band. This step has been followed by a graph quantification of the constructed networks. Our results showed that AD networks are characterized by reduced global and local connectivity compared to healthy controls in alpha1 frequency band. In addition, a significant correlation between the clinical test (MMSE score) and the EEG graph metrics was obtained.

## II. MATERIALS AND METHODS

The full pipeline of the study is illustrated in Figure 1.

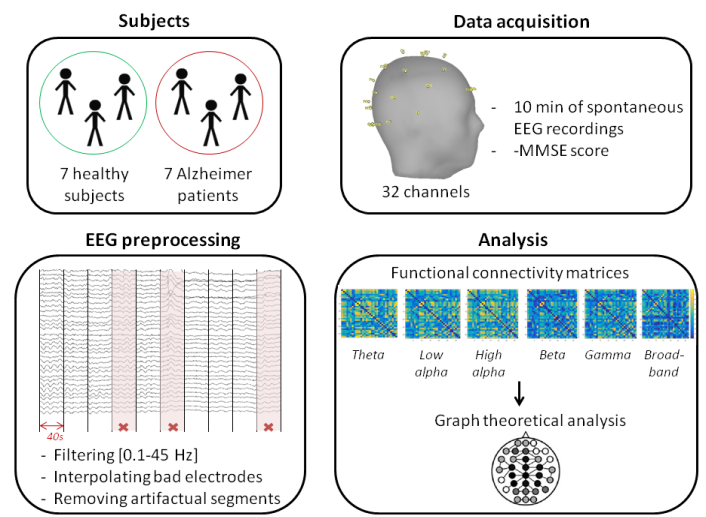

**Figure 1. The full pipeline of the study**

### A. Participants

Seven healthy controls (5 males and 2 females, age 64–74 y) and seven patients diagnosed with AD (4 females and 3 males, age 69–81 y) participated in this study. AD patients were recruited at the memory clinic of Al-Ajaza Hospital and Mazloum Hospital, Tripoli, Lebanon. None of the patients was suffering from another neurological disease. Controls were recruited from the local community, and from Al-Ajaza Hospital. Prior to their participation, all subjects have provided informed consent. The study was approved by the local institutional review boards (CE-EDST-3-2017).

## B. Data acquisition

EEG signals were recorded using a 32-channel EEG system (Twente Medical Systems International (TMSi), Porti system). Electrodes were placed in accordance with the revised international 10/20 system [11] at Fpz, Fp1, Fp2, Fz, F3, F4, FC1, FC2, FC5, FC6, C3, C4, Cz, CP1, CP2, CP5, CP6, P3, P4, P7, P8, Pz, PO3, PO4, PO7, PO8, POz, O1, O2, Oz, T7 and T8. The wristband was used as ground. EEG signals were sampled at 512 Hz, bandpass filtered within 0.1–80 Hz, and stored in GDF format.

Subjects were asked to relax with their eyes-closed without falling asleep for 10 minutes. We also screened their cognitive score using the mini-mental state examination (MMSE) [12].

## C. EEG preprocessing

EEG signals were filtered between 0.1 and 45 Hz, and segmented into 40s epochs. Segments with amplitude ±90μV were considered as artifactual and were eventually discarded from the analysis after visual inspection. For some cases, the electrodes detected as noisy were interpolated. For each participant, we selected the maximum number of non-artifactual segments.

## D. Functional connectivity analysis

The functional connectivity matrices representing the estimates of interaction between all pairs of EEG channels were computed in the different frequency bands ($\theta$(4–8 Hz), $\alpha1$(8–10 Hz), $\alpha2$(10–13 Hz), $\beta$(13–30 Hz), $\gamma$(30–45 Hz), and broad band (0.2-45 Hz)). Here, the connectivity was assessed using the phase synchronization (PS) method as it is less susceptible to the effects of artifacts [13] and was successfully used in several scalp-level EEG connectivity analyses [14], [15] such as Brain Computer Interfaces. In particular, we used the phase locking value (PLV) proposed in [16] and defined as:

$$PLV(t) = \left| \frac{1}{\delta} \int_{t-\delta/2}^{t+\delta/2} exp(j(\varphi_y(t) - \varphi_x(t))d\tau \right|$$

where $\varphi_y(t)$ and $\varphi_x(t)$ are the unwrapped phases of the signals $x$ and $y$ at time $t$. The Hilbert transform was used to extract the instantaneous phase of each signal. $\delta$ denotes the size of the window in which PLV is calculated. The connectivity matrices were thresholded by using a 10% proportional threshold. For each subject, we obtained the frequency-specific networks, in which, nodes signify the 32 electrodes and edges represent the connectivity value between channels (PLV).

## E. Graph metrics extraction

In order to study the difference between healthy and AD networks, graph theoretical analysis was used [17]. Here we used three graph metrics:

1- Clustering coefficient: The clustering coefficient of a node quantifies how close its neighbors are to being a clique [18]. This parameter is one of the most elementary measures of local segregation (i.e. the extent to which the network is organized into specialized functional regions). [17] The average of the clustering coefficients for each individual node is the clustering coefficient of the graph.
2- Global efficiency: The global efficiency is computed as the average inverse shortest path length [19]. A short path length indicates that, on average, each node can be reached from any other node along a path composed of only a few edges [6]. The global efficiency measures the global integration of a network (i.e the capacity of the network to combine information from many distributed areas).
3- Strength: It evaluates the importance of a node within a network. This measure is defined as the sum of the edges' weights linked to the node [20].

## III. RESULTS

First, for each subject, the connectivity matrices of all segments were averaged, and then the different graph metrics (mean PLV value, average clustering, and global efficiency) were extracted. Figure 2.A illustrates the difference between healthy and AD subjects at each frequency band. The mean functional connectivity as well as the average clustering and the global efficiency in the AD group were decreased in the lower alpha band compared to controls ($p <0.001$, Wilcoxon test) (Figure 2.A). As illustrated, no between group differences were obtained for theta, alpha2, beta, and gamma frequency bands ($p>0.01$).

Second, we assessed the correlation between the cognitive state of each subject (MMSE score) and the extracted graph measures (Figure 2.B). Results show a positive significant correlation for the mean connectivity (R=0.81; $p<0.001$), the average clustering (R=0.82, $p<0.001$), and the global efficiency (R= 0.81, $p<0.001$).

We were also interested in identifying the nodes that show decreased activity in AD patients compared to controls, in terms of strength and clustering coefficient. For this aim, we concatenated the metrics distributions of each node across subjects at alpha1 frequency band. Then, we computed the difference between the two groups on each individual node using Wilcoxon test. Results depict that the nodes that show significantly reduced functionality in the AD group ($p<0.001$) with regard to the clustering coefficient are: Oz, Fp1, Fp6, FC6, FC1, POz. According to the strength, the disrupted nodes are: Oz, O2, C3, C4, Fp2, Fpz, Fp1, FC5, FC1, FC2, F4, POz, PO8, Fz.

## IV. DISCUSSION

Results show that AD alters the global and local properties of the scalp-level brain network. The decreased functional synchronization obtained in AD patients compared to healthy controls agrees with previous finding [8], [21]–[23]. Similarly, the decreased global efficiency of the AD network is consistent with [7], [9]. Results revealed an association of the alterations in brain network organization with cognitive deficits (MMSE clinical test) in AD. This was also observed in several studies [7]–[9], [21], [22]. However, the reduced segregation (low clustering coefficient) associated mainly to the frontal regions is in contradiction with multiple studies as reported in this review [23]. A possible explanation of this difference is the use of different neuroimaging techniques (with different spatial resolutions) and different methodologies (connectivity measure, graph measure).

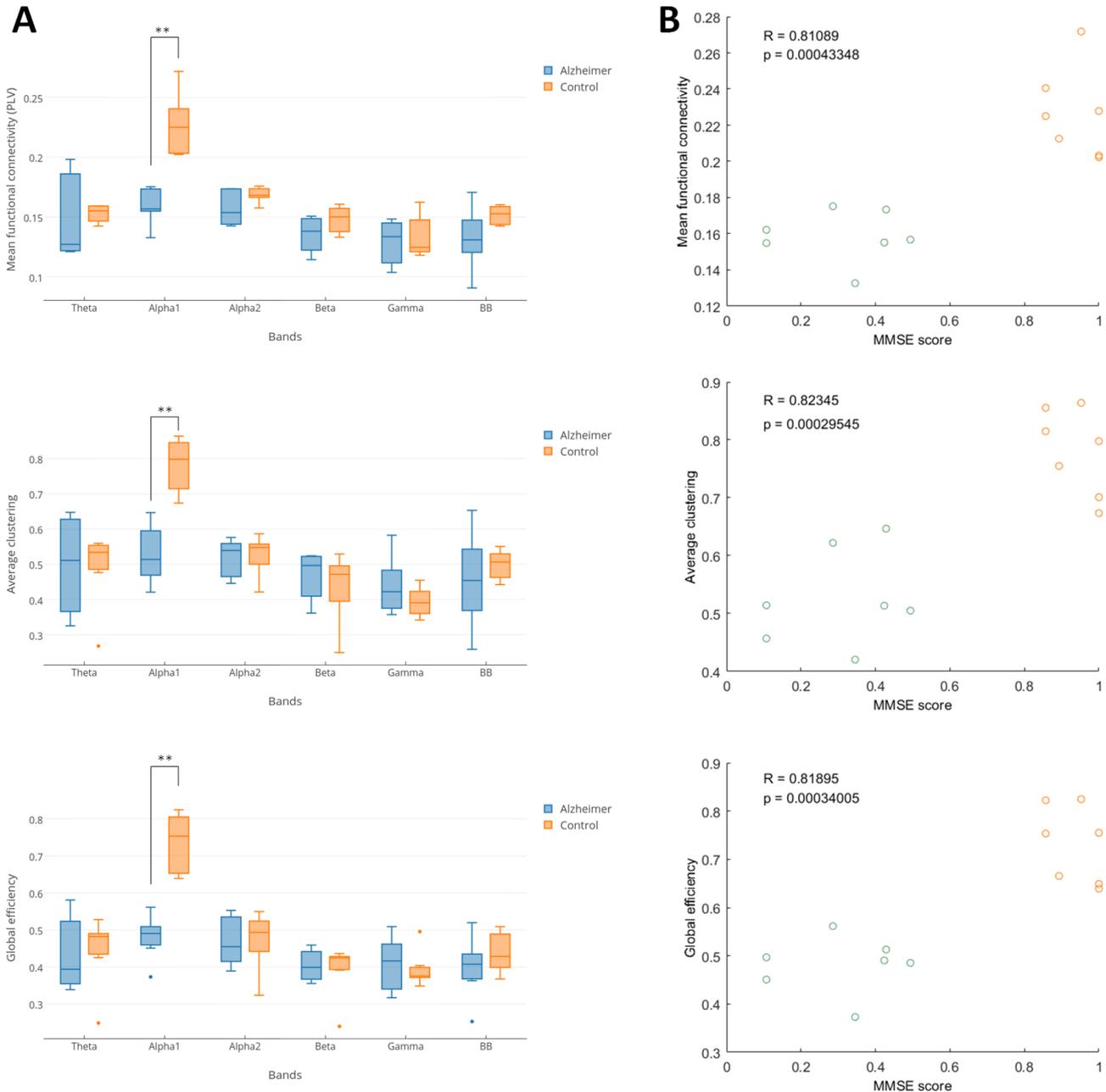

Figure 2. A) The difference between AD patients and healthy networks in terms of 1) mean functional connectivity, 2) average clustering and global efficiency for the six frequency bands and B) The correlations between the MMSE score and the graph metrics in the alpha1 band.

This preliminary study has some limitations that should be addressed. First, the functional connectivity was computed on the scalp (electrodes level) that could be generally corrupted by the volume conduction problem. To overcome this problem and to improve the spatial resolution of the networks, 'EEG source connectivity' method is an emerging technique used in cognition [24]–[26] and brain disorders [27][28]. However, in the current study, we were cautious to reconstruct the brain sources because of the relatively low number of available channels (32 electrodes) which may affect the accuracy of the results [24]. Therefore, a further work is to take into account this technical issue and adapt the EEG source connectivity method in order to accurately identify the cortical networks using relatively low number of channels. Second, we used a proportional threshold to remove weak connections of the functional connectivity matrices. This threshold was used to ensure equal density across patients and control subjects. Nevertheless, one should test the reproducibility of results by applying other proportional thresholds. Finally, the analysis should be extended on a larger dataset to obtain consistent results across subjects.

V. CONCLUSION

In this paper, we used EEG scalp connectivity method to explore the alterations in the networks organization of AD patients. The synchronization was assessed using the PLV measure, and the networks were characterized in terms of average clustering, global efficiency and mean functional connectivity. Our results show a correlation between these measures and the MMSE score in the lower alpha band. This preliminary study opens insights on the capacity of EEG-based functional connectivity method to identify the pathological brain networks during rest.

ACKNOWLEDGMENT

This work has been funded with support from the National Council for Scientific Research in Lebanon.